\documentclass[groupedaddress, superscriptaddress, twocolumn, amsmath, amssymb, aps]{revtex4-2}
\usepackage[english]{babel}
\usepackage{latexsym}
\usepackage{graphics}

\usepackage{amsmath}
\usepackage{amssymb}
\usepackage{graphicx}
\usepackage{physics}
\usepackage{xcolor, tikz}
\usepackage[colorlinks=true, citecolor=blue]{hyperref}
\usetikzlibrary{quantikz2}

\begin{document}

\title{Qudit-native simulation of the Potts model}

\author{Maksim A. Gavreev}
\email{m.gavreev@rqc.ru}
\affiliation{National University of Science and Technology ``MISIS'', Moscow 119049, Russia}

\author{Evgeniy O. Kiktenko}
\affiliation{National University of Science and Technology ``MISIS'', Moscow 119049, Russia}

\author{Aleksey K. Fedorov}
\affiliation{National University of Science and Technology ``MISIS'', Moscow 119049, Russia}

\author{Anastasiia S. Nikolaeva}
\affiliation{National University of Science and Technology ``MISIS'', Moscow 119049, Russia}

\begin{abstract}
Simulating entangled, many-body quantum systems is notoriously hard, especially in the case of high-dimensional nature of underlying physical objects.
In this work, we propose an approach for simulating the Potts model based on the Suzuki-Trotter decomposition that we construct for qudit systems.
Specifically, we introduce two qudit-native decomposition schemes: (i) the first utilizes Mølmer–Sørensen gate and
 additional local levels to encode the Potts interactions , while (ii) the second employs an light-shift gate thatnaturally fits qudit architectures. 
These decompositions enable a direct and efficient mapping of the Potts model dynamics into hardware-efficient qudit gate sequences for trapped-ion platform.
Furthermore, we demonstrate the use of a Suzuki--Trotter approximation with our evolution-into-gates framework, for detecting the dynamical quantum phase transition. Our results establish a pathway toward qudit-based digital quantum simulation of many-body models and provide a new perspective on probing nonanalytic behavior in high-dimensional quantum many-body models.
\end{abstract}

\maketitle

\section{Introduction}\label{Sec:Intro}

Quantum computational devices open the way to explore entangled, many-body quantum states that are believed to be hard for classical analysis in the controllable manner~\cite{Preskill2012}.
The most well-developed digital model of quantum computing follows along the lines of classical information theory~\cite{Brassard1998}, 
so that the information units in the quantum domain are qubits that are quantum counterparts of classical information bits. 
However, underlying physical objects in physical systems that are used for quantum computing~\cite{Schoelkopf2013,Gambetta2017,Martinis2018} are essentially multilevel~\cite{Fedorov2018}, 
e.g. neutral atoms and trapped ions with a rich structure of energy levels
(so that their use as qubits requires idealization). 
As soon as one of the key potential applications for quantum computing with the simulation of complex (entangled and many-body) quantum systems, 
then this feature of ability to manipulate multilevel states can used for efficient simulation of quantum models consists of high-dimensional objects, such as high-order spins~\cite{RICO2018466}.

We note that the use of multilevel systems, also known as qudits, in quantum computing is of general interest also in the context of a more efficient implementation of quantum algorithms (for a review, see Ref.~\cite{RevModPhys.97.021003, Nikolaeva2021epj})
Moreover, one can admit is significant progress in the realization of qudit-based processors with trapped ions~\cite{Zalivako2024qb16, Ringbauer2022}, superconducting systems~\cite{Hill2021, Goss2023Toffoli}, and quantum light~\cite{Wang2022}.
Several experiments with the use of qudit processors for quantum simulations~\cite{PhysRevA.109.032619, Meth2025, PhysRevA.110.062602, Chicco2024, rrmd-fqx6}, for example in the case of high-energy physics models, have been performed~\cite{PRXQuantum.4.027001, 2D_electrodynamics}. 

A prominent example of a quantum many-body model with a wide range of applications is the Potts model~\cite{CHENG201788, SZABO20161, LI20211}, 
which is a generalization of the famous quantum Ising model (the Potts model is also related to the so-called  chiral clock model~\cite{PhysRevA.98.023614, PhysRevB.109.235104, PhysRevB.98.245104}). 
The key feature of the Potts model is the ability to capture more reach structures in quantum phase transitions \cite{Bernien2017, Keesling2019, Samajdar2018}, critical phenomena \cite{Keesling2019}, exotic quasiparticle excitations (e.g., mesonic and baryonic) \cite{liu2020}, 
properties of integrable lattice models \cite{martin1991potts}, and entanglement structures~\cite{Lotkov2022}.
However, as in the case of the Ising model starting from a certain number of spins and their entanglement structure, the Potts model starts to be intractable with classical computational devices 
(we note that various approximate approaches, such as Monte Carlo~\cite{PAN1992773, Wen_2011} and projected entangled-pair state (PEPS)~\cite{SMIERZCHALSKI2025102257} simulations have been performed).
We also would like to mention that transitions into complex $\mathbb{Z}_n$-ordered phases, where excitations are evenly separated by $n>2$ sites, 
have been explored in experiments with programmable Rydberg quantum simulator~\cite{Keesling2019};
the possibility to explore the critical properties of the three-state Potts model with fine-tuned pulses has been also mentioned. 


In this work, we propose the way to study the Potts model with the use of the Suzuki-Trotter (ST) decomposition that we construct for qudit systems.
For this purpose, we introduce two new qudit-native decomposition schemes. 
The first scheme is based on the light-shift (LS) gate that naturally fits qudit architectures \cite{Hrmo2023}.
The second approach utilizes additional local levels to encode the Potts interactions.
These decompositions allow one to map the Potts model dynamics into hardware-efficient gate sequences. 
As a target qudit hardware platform we consider trapped-ion platform, which is widely used for qudit-based simulations \cite{Meth2025, Kazmina2024demonstration, joshi2025} and algorithm  realizations \cite{Zalivako2024qb16, Low2025_25levels, Nikolaeva2025prl}. We also conduct numerical experiments to demonstrate how gate decompositions can be used within the Suzuki-Trotter framework for dynamics simulation. We demonstrate how the presented decompositions can be used to observe a dynamic phase transition in the Potts model.

The paper is organized as follows. In Sec.~\ref{Sec:Model}, we introduce the Potts model in its canonical formulation. The Suzuki–Trotter (ST) decomposition scheme is discussed in Sec.~\ref{Sec:Trotter}. In Sec.~\ref{Sec:Decompositions}, we present the qudit-native decompositions of the single- and two-qudit gates arising within the ST framework. Numerical simulation results demonstrating the dynamical quantum phase transition (DQPT) in the Potts model are provided in Sec.~\ref{Sec:DQPT}. We summarize our findings in Sec.~\ref{Sec:Conclusion}.

\section{Quantum Potts model}\label{Sec:Model}
The quantum Potts model generalizes the well-known transverse-field Ising model by extending the local Hilbert space from two to $q$ internal states. Each site of the system is thus described by a $q$-dimensional qudit, and the model captures a wider range of symmetry-breaking and critical phenomena arising from the $\mathbb{Z}_q$ discrete symmetry group. The Hamiltonian of the 1D quantum Potts chain can be written as
\begin{eqnarray}\label{eq:hamiltonian}
&&H = H^{I} + H^{L},\\
&& H^{I} = - J \sum_{\left< n, n'\right>} H^{I}_{n, n'}, \quad H^{L} = - g \sum_{n} H_{n}^{L}\\
&&H^{I}_{n, n'} = \sum_{k=1}^{q-1} \Omega_{n}^{k}\Omega_{n'}^{q-k} \quad H^{L}_{n} = \sum_{k=1}^{q-1} \Gamma_{n}^{k},
\end{eqnarray}
where $\Omega$ and $\Gamma$ are clock and shift operators respectively.
\begin{equation}
\Omega  = 
\begin{pmatrix}
1 &&& \\
&\omega&&\\
&&\ddots&\\
&&&\omega^{q-1}\\
\end{pmatrix},
\quad
\Gamma = \begin{pmatrix}
 & \mathbb{I}_{q-1} \\
1 & \\
\end{pmatrix},
\end{equation}
and $\omega=e^{2\pi  i/q}$ $q$-th root of unity and $I_{q-1}$ is the identity matrix of dimension $q-1$. 

Clock and shift operators can be considered as high-dimensional generalizations of qubit $Z, X$ operators. In the special case $q=2$, $\Omega$ and $\Gamma$ reduce to the standard Pauli matrices $\sigma_z$ and $\sigma_x$, respectively, and the Hamiltonian becomes equivalent to the transverse-field Ising model. 

The interplay between the interaction term, which energetically favors alignment of neighboring qudits, and the transverse mixing term, which induces transitions between local states, gives rise to a rich dynamical and equilibrium phase structure. For $|g| \ll |J|$, the system exhibits a ferromagnetic phase characterized by ordered configurations with all qudits occupying the same internal state. In the opposite limit $|g| \gg |J|$, the eigenstates become delocalized superpositions of all local levels, forming a disordered, quantum-paramagnetic phase.

From the perspective of quantum simulation, the Potts model represents an ideal benchmark for studying the effect of dimensionality in local Hilbert spaces and for exploring critical dynamics in multi-level quantum systems. The presence of non-commuting interaction and mixing terms makes the model analytically intractable in most regimes, especially out of equilibrium.

\section{Suzuki--Trotter Decomposition}\label{Sec:Trotter}

Digital quantum simulation provides a framework for approximating the continuous-time dynamics of quantum many-body systems by a discrete sequence of unitary gates that act on local subsystems~\cite{Lloyd1996}. The goal is to express the total time-evolution operator generated by a Hamiltonian $H$ as a product of unitary operators that each act on a small subset of degrees of freedom.

Consider a general Hamiltonian Eq.~\ref{eq:hamiltonian} written as a sum of interaction and local mixing terms where operators $H^{I}, H^{L}$ typically do not commute with one another $[H^{I}, H^{L}] \neq 0$. The first-order Lie--Trotter approximation replaces the global exponential with a product of local exponentials evaluated over a small time step $\tau = t/m$:
\begin{equation}
U(t) = \left[\exp(-i \tau H^{I}) \exp(-i \tau H^{L})\right]^{m} + \mathcal{O}(\tau^2/m).
\end{equation}
Higher-order Suzuki–Trotter formulas symmetrize exponentials to suppress this error to higher powers of $\tau$ at the expense of additional gate operations. This trade-off between accuracy and circuit depth plays a central role in practical implementations of digital quantum simulation. For the $q$-state Potts Hamiltonian, applying the second-order Suzuki--Trotter formula to this Hamiltonian gives
\begin{equation}
U(t) =
\left[
U^{L}(\tau/2)U^{I}(\tau)U^{L}(\tau/2)
\right]^{m} + \mathcal{O}(\tau^3/m^2), 
\end{equation}
where 
\begin{equation}
U^{L}(\tau) = \exp(-i\tau H^{L}), \quad U^{I}(\tau) = \exp(-i\tau H^{I}).
\end{equation}
Since each hermitian operator under exponential is a sum of commuting terms we can expand each unitary into product 
\begin{align}
&U^{I}(\tau) = \prod_{\left<n,n'\right>} U^{I}_{n,n'}(\tau), \quad U^{L}(\tau) = \prod_{n} U^{L}_{n}(\tau),\\
&U^{I}_{n,n'}(\tau) = \exp(i \tau J H^{I}_{n,n'}),~U^{L}_{n}(\tau) = \exp(i \tau g H_{n}^{L}),
\end{align}
where each exponential factor can be identified as an elementary evolution gate acting either on a pair of qudits or on a single qudit. 

The total Trotterized evolution therefore consists of $m$ sequential layers, each composed of parallel applications of $\exp(i \tau J H^{I}_{n,n'})$ followed by $\exp(i \tau g H_{n}^{L})$. In qudit-based architectures such as trapped-ion or Rydberg-atom platforms, these gates correspond to experimentally realizable multi-level interactions and local drives. The Suzuki--Trotter decomposition thus bridges the gap between the continuous-time dynamics of the Potts Hamiltonian and a discrete gate-based realization, providing a mathematically controlled approximation whose fidelity can be systematically improved by increasing the approximation order or decreasing the step size $\tau$.


\section{Gate decompositions}\label{Sec:Decompositions}
\subsection{Single-qudit gate decomposition}\label{SubSec:1Q}

Within the Suzuki--Trotter framework, the non-diagonal part of the Potts Hamiltonian, $H^{L}$ generates local single-qudit transitions between internal qudit levels. The generalized shift operator $\Gamma$ acts as
$\Gamma |m\rangle = |(m+1)\bmod q\rangle$. and satisfies the commutation relation $\Gamma \Omega = \omega\, \Omega \Gamma$ with $\omega = e^{2\pi i/q}$. 

The operator $\Gamma$ is diagonal in the Fourier basis. 
\begin{equation}
\Gamma= F^{\dagger}_{q}\, \Omega^{k}\, F_{q},
\end{equation}
where $F_{q}$ is the discrete Fourier transform in $q$-dimensional basis. Summing over $k$ yields
\begin{equation}
H_{n}^{L} 
    = F^{\dagger}_{q} \left( \sum_{k=1}^{q-1} \Omega^{k} \right) F_{q}
    = F_{q}^{\dagger} D F_{q},
\end{equation}
with $D = \mathrm{diag}(q-1,-1,\ldots,-1)$. Substituting this diagonalization into the mixer evolution operator gives
\begin{equation}
U^{L}_{n}(\tau) = F^{\dagger}_{q} \exp \left( i g\tau\, D \right) F_{q}
\end{equation}
We note that the diagonal phase gate can be implemented virtually~\cite{McKay2017}. Thus, the entire cost of realizing $U^{L}_{n}(\tau)$ is determined by the cost of implementing two discrete Fourier transforms.

A $q$-dimensional qudit Fourier transform $F_{q}$ can be decomposed into a sequence of native two-level rotations of the form $R^{ab}(\theta,\phi) = \exp(-i\theta/2 [\cos(\phi) \sigma_{x}^{ab} + \sin(\phi) \sigma_{y}^{ab}])$ acting on pairs of basis levels, as shown in~\cite{drozhzhin2025}. A standard construction requires at most $\frac{q(q-1)}{2}$
two-level rotations. Since the mixer gate requires two Fourier transforms, the total decomposition cost is $q(q-1)$. Because the diagonal phase shift is virtual, this bound is tight, the mixer gate requires exactly the same cost as two Fourier transforms.


\subsection{Two-qudit gate decompositions}\label{SubSec:2Q}
\subsubsection{LS-gate-based decomposition}
We now focus on the decomposition of the two-qudit interaction term arising in the Potts Hamiltonian. The central object of interest is the interaction operator $H^{I}_{n,n'}=\sum_{k=1}^{q-1} \Omega^k \Omega^{q-k}$. This operator describes the coupling between two qudits mediated through their “clock” operators, and it plays the role of an interaction term. To understand its structure, let us evaluate the matrix elements of $H^{I}_{n,n'}$ in the product basis $\{\ket{s, s'}\}$, for $s, s' \in \{0, 1, \dots, q-1\}$ for fixed $n,n'$. We find 
\begin{equation}
\langle s, s'| H^{I}_{n,n'}|s, s'\rangle=\sum_{k=1}^{q-1} \omega^{k(s - s')}.
\end{equation}
The summation above is a discrete Fourier sum over the phase difference between the two levels $s$ and $s'$. It evaluates to $q$ when $s=s'$ and to zero otherwise, since
\begin{equation}
H^{I}_{n,n'}\ket{s, s'} = \left\{
\begin{array}{ll}
(q-1) \ket{s, s'}, \quad s = s' \\
-1, \quad s \neq s'.
\end{array}\right.
\end{equation}
Hence, the operator $H^{I}_{n,n'}$ acts as a projector onto the subspace where both qudits occupy the same level. Explicitly, we can write
\begin{equation}
H^{I}_{n,n'}=q \sum_{s=0}^{q-1}|s, s\rangle\langle s, s| - \mathbb{I}=q \Pi_{\text {same}} - \mathbb{I},
\end{equation}
where $\Pi_{\text {same }}$ is the projector onto the symmetric subspace of equal-level states. The constant shift by $- \mathbb{I}$ contributes only a global phase in the time evolution and can therefore be neglected. The two-qudit time-evolution operator generated by $H^{I}_{n,n'}$ during a time interval $t$ reads
\begin{equation}
U^{I}_{n,n'}(\tau) = \exp(i \tau q J \Pi_{\text{same}}),
\end{equation}
where $J$ is the interaction strength. The exponential form shows that $U(t)$ acts trivially on all states with $s \neq s'$, while states with $s = s'$ acquire a phase shift proportional to $\tau qJ$. The gate therefore applies a conditional phase depending on whether the two qudits are in the same state. 

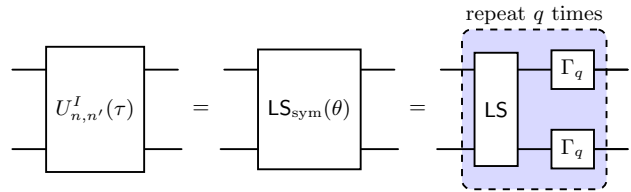
\begin{figure}[t]
    \centering
\resizebox{.98\linewidth}{!}{
\begin{quantikz}
& \gate[2]{U^{I}_{n,n'}(\tau)}  &\midstick[2, brackets=none]{=}  & \gate[2]{{\sf LS}_{\rm sym}(\theta)}  &\midstick[2, brackets=none]{=} &   \gate[2]{\sf LS} \gategroup[wires=2,steps=2,background, style={dashed,rounded corners,fill=blue!15, inner xsep=2pt, opacity=20}]{repeat $q$ times} & \gate{\Gamma_{q}} & \qw \\
&&&& &                             & \gate{\Gamma_{q}}         & \qw
\end{quantikz}}
    \caption{Two-qudit time-evolution operator implementation using light shift (LS) gate.}
    \label{fig:circuit-ls}
\end{figure}

Such a phase pattern corresponds exactly to the experimentally realized symmetric light-shift (LS) gate introduced for trapped-ion qudit platform in~\cite{Hrmo2023} and obtained by symmetrizing its non-symmetric version ${\sf LS}$ (Fig.~\ref{fig:circuit-ls}). This gate acts as
\begin{equation}
\mathsf{LS}_{\rm sym}(\theta):\left\{
\begin{array}{ll}
\ket{s, s} \mapsto \ket{s, s} \\
\ket{s, s'} \mapsto e^{i\theta} \ket{s, s'}, & s \neq s',
\end{array}\right.
\end{equation}
and can equivalently be written as
\begin{equation}
\mathsf{LS}_{\rm sym}(\theta) = e^{i\theta} \exp(i \theta \Pi_{\text{same}}).
\end{equation}
Comparing this with the expression for $U^{I}_{n,n'}(t)$ above, we immediately identify the correspondence $\theta= \tau qJ$ (for LS gate parameters tuning see~\cite{Hrmo2023} Supplementary material). Hence, the LS gate provides a native realization of the two-qudit evolution operator required by the Potts model requiring $\mathcal{O}(q^2)$ single qudit gates and $\mathcal{O}(q)$ two-qudit gates. Its diagonal structure ensures that the interaction is purely phase-based and does not mix different computational basis states, making it particularly suitable for Trotterized simulation schemes where diagonal and off-diagonal terms alternate.

\subsubsection{Decomposition based on an additional level}

\begin{figure*}[ht!]
    \centering
\resizebox{.98\linewidth}{!}{
\begin{quantikz}
& \gate[2]{U^{I}_{n,n'}(\tau)}  &\midstick[2, brackets=none]{=}  
& \gate{V^{\dagger,0}_{n}} 
\gategroup[wires=2,steps=3,background, style={dashed,rounded corners,fill=blue!15, inner xsep=2pt, opacity=20}]{}  
& \gate[2]{{\sf MS}^{0}_{n,n'}(\theta)}  
& \gate{V^{0}_{n}} 
& &\midstick[2, brackets=none]{\dots}  
& \gate{V^{\dagger, q-1}_{n}} 
\gategroup[wires=2,steps=3,background, style={dashed,rounded corners,fill=blue!15, inner xsep=2pt, opacity=20}]{}  
& \gate[2]{{\sf MS}^{q-1}_{n,n'}(\theta)}  
& \gate{V^{q-1}_{n}} 
&\\
&       &&\gate{V^{\dagger, 0}_{n'}}&&  \gate{V^{0}_{n'}}                    &  &    &\gate{V^{\dagger, q-1}_{n'}}&&  \gate{V^{q-1}_{n'}}                    &      
\end{quantikz}}
    \caption{Two-qudit time-evolution operator implementation using ancilla level.}
    \label{fig:circuit-ms}
\end{figure*}
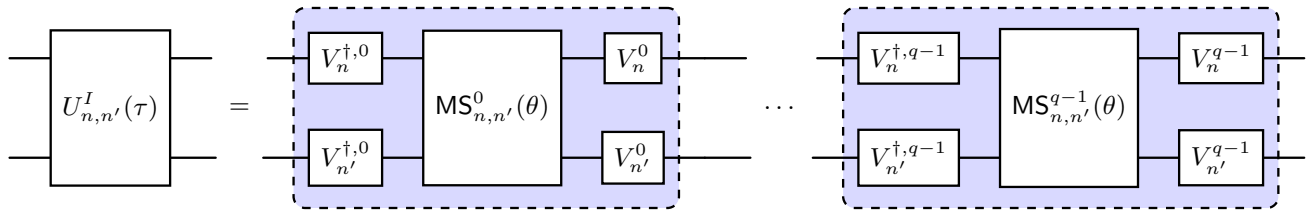

In some architectures, direct implementation of the LS-type interaction may not be available. An alternative route is to exploit an extended Hilbert space containing an additional, auxiliary level. This additional level enables the realization of the projector $\Pi_{{\rm same}}$ using pairwise interactions between effective two-level systems.

Let each physical qudit possess one ancilla level, denoted $\ket{q}$, in addition to levels $\{\ket{0}, \ket{1}, \ldots, \ket{q-1}\}$. The full local Hilbert space is therefore $\mathcal{H} = {\rm span} \{\ket{0}, \ket{1}, \ldots, \ket{q-1}, \ket{q}\}$. Within each two-level subspace spanned by $\{\ket{k}, \ket{q}\}$ we define Pauli operators 
\begin{equation}
\sigma_{z}^{k} = \ketbra{k}{k} - \ketbra{q}{q}, \qquad
\sigma_{x}^{k} = \ketbra{k}{q} + \ketbra{q}{k}.
\end{equation}
These operators satisfy the same algebra as standard Pauli matrices in their respective subspaces. 
Using these definitions, we introduce the operator
\begin{equation}
\Pi_{n, n'} = \sum_{k=0}^{q-1} \sigma_{z,n}^{k} \sigma_{z,n'}^{k},
\end{equation}
which acts jointly on two qudits $n$ and $n'$. Lets consider its action on the product basis states $\ket{s, s'}$ within the logical subspace $s, s' \in \{0, 1, \ldots, q-1\}$ for fixed $n,n'$. For each pair $(s, s')$
\begin{equation}
\begin{aligned}
    \sigma_{z,n}^{k} \sigma_{z,n'}^{k}\ket{s, s'} = (\delta_{s,k} - \delta_{s,q})(\delta_{s',k} - \delta_{s',q}) \ket{s, s'} = \\ 
    = \delta_{s,k}\delta_{s',k}\ket{s, s'},
\end{aligned}
\end{equation}
since $\delta_{s,q} = 0$ and $\delta_{s',q} = 0$ in the logical subspace. Summing over $k$ gives $\Pi_{n,n'} \ket{s, s'} = \delta_{s, s'}\ket{s, s'}$. Therefore, within the logical subspace, the operator $\Pi_{n,n'}$ acts identically to the projector $\Pi_{{\rm same}}$. This observation allows us to replace $\Pi_{{\rm same}}$ by $\Pi_{n,n'}$ in the desired evolution operator. The corresponding two-qudit unitary is then
\begin{equation}\label{eq:anc-decomp}
U^{I}_{n,n'}(\tau) = \exp(i \theta\Pi_{n,n'}) = \prod_{k=0}^{q-1} \exp(i \theta \sigma_{z,n}^{k}\sigma_{z,n'}^{k}),
\end{equation}
for $\theta = \tau q J$. The factorization in Eq.~\eqref{eq:anc-decomp} follows from the commutativity of the individual $\sigma_{z,n}^{k}\sigma_{z,n'}^{k}$ terms acting on distinct subspaces. Each term represents a two-level conditional phase gate between the pair of subspaces $\{\ket{k}, \ket{q}\}_{i} \otimes \{\ket{k}, \ket{q}\}_{j}$. To express these interactions in a more experimentally accessible form, it is convenient to rotate each subspace so that the coupling appears in the $\sigma_{x}$-basis. Defining
\begin{equation}
V^{k}_{n} = \exp \left(-i \frac{\pi}{4}\sigma_{y}^{k}\right),
\end{equation}
which maps $\sigma_{z}^{k} \mapsto \sigma_{x}^{k}$, we can rewrite the evolution as
\begin{equation}
U^{I}_{n,n'}(\tau) = \prod_{k=0}^{q-1} (V_{n}^{k} \otimes V_{n'}^{k}) {\mathsf{MS}}_{n,n'}^{k}(\theta) (V_{n}^{\dagger, k} \otimes V_{n'}^{\dagger, k}),
\end{equation}
where
\begin{equation}
{\mathsf{MS}}_{n,n'}^{k}(\theta) = \exp(i \theta \sigma_{x, n}^{k}\sigma_{x, n'}^{k})
\end{equation}
denotes the Mølmer–Sørensen (MS) gate acting within the two-level subspace $\{\ket{k}, \ket{q}\}$, see Fig.~\ref{fig:circuit-ms}. The decomposition above shows that the entire two-qudit evolution $U^{I}_{n,n'}(\tau)$ can be realized as a sequence of $\mathcal{O}(q)$ independent MS gates and $\mathcal{O}(q)$ single qudit gates, each acting on a distinct two-level manifold and conjugated by local rotations. In this way, the presence of a single auxiliary level per qudit suffices to emulate the multilevel interaction structure of the Potts Hamiltonian using only single-qudit rotations and widely used Mølmer–Sørensen two-qudit gates~\cite{Ringbauer2022, Zalivako2024qb16}.

Conceptually, this construction demonstrates that even complex multiqudit projectors such as $\Pi_{{\rm same}}$ can be synthesized from simple pairwise couplings when an extended Hilbert space is available. It also highlights the modular nature of qudit control: by combining local subspace rotations and entangling gates, one can reproduce higher-dimensional interactions without requiring direct multilevel entanglement. This decomposition is therefore particularly suitable for scalable simulation architectures, as it enables implementation of the Potts-type two-qudit interactions entirely within the standard MS gate framework.

\section{Qudit simulation of the Potts model: dynamical quantum phase transition}\label{Sec:DQPT}

\begin{figure}[t!]
    \centering 
    \includegraphics[width=0.48\textwidth]{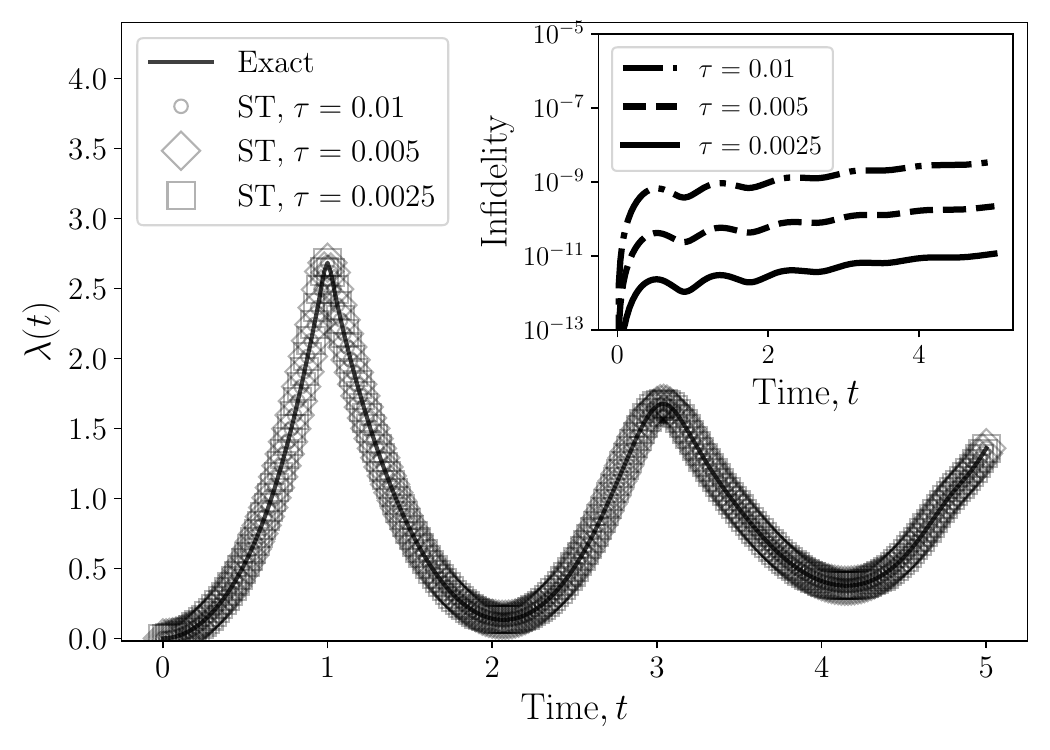} 
    \vspace{-20pt}
    \caption{Rate function for the simulated dynamics with the qutrit Potts Hamiltonian. The inset shows the infidelity caused by Trotterization error.}
    \label{fig:dqpt} 
\end{figure}

To assess the accuracy and physical relevance of the presented decompositions, we perform a numerical simulation of the nonequilibrium dynamics governed by the qutrit ($q=3$) Potts Hamiltonian in a finite chain of $N=6$ qudits. The evolution is initialized in the ferromagnetic ground state of the interaction Hamiltonian, $\ket{\psi_0} = \bigotimes_{i=1}^{N}\ket{0}_{i}$, and the subsequent unitary dynamics is driven by a sudden quench of the transverse parameter $g$, which activates the non-diagonal mixing term. The resulting evolution operator is applied iteratively for $N_t = t / \tau$ time steps, corresponding to total evolution time $t$. The system parameters are $J=1/4, g=1$.

A key observable characterizing the dynamical behavior is the return probability, or Loschmidt echo, defined as
\begin{equation}
\mathcal{L}(t) = |\langle \psi_0 | \exp({-itH}) | \psi_0 \rangle|^2,
\end{equation}
which quantifies the overlap between the initial and time-evolved states. To make the analysis analogous to thermodynamic phase transitions, one introduces the rate function,
\begin{equation}
\lambda(t) = -\frac{1}{N} \log \mathcal{L}(t),
\end{equation}
which plays the role of a dynamical free-energy density. Nonanalytic behavior or sharp peaks in $\lambda(t)$ are interpreted as signatures of a dynamical quantum phase transition, reflecting critical changes in the temporal structure of the wave function under unitary evolution.

In our simulations, we compute $\lambda(t)$ using both the exact diagonalization of the full Hamiltonian and the Suzuki--Trotter decomposition with time step $\tau$. The comparison demonstrates that even for moderate Trotter step sizes, the numerical approximation captures the essential dynamical features of the system. The characteristic cusps in the rate function appear at the same critical times as in the exact solution, indicating that the second-order Trotterized dynamics reproduces the DQPT signatures with high fidelity.

The results are summarized in Fig.~\ref{fig:dqpt}, where the normalized logarithm of the return probability is plotted as a function of evolution time. The agreement between the exact and Trotterized results confirms the robustness of the decomposition scheme for simulating real-time dynamics in multi-level spin systems. 

Physically, the appearance of nonanalyticities in $\lambda(t)$ corresponds to critical points in the quantum state's evolution, where the system undergoes coherent population transfer among macroscopically distinct configurations. In the $q=3$ Potts chain, this behavior reflects oscillatory competition between ordered domains and disordered superpositions induced by the transverse mixing. The observation of DQPTs in the simulated dynamics demonstrates that the qudit-based Suzuki--Trotter approach provides an efficient and accurate tool for studying non-equilibrium quantum phenomena beyond the qubit limit.

These results validate the applicability of the Trotterized framework for future digital emulations of higher-dimensional Potts models and for benchmarking near-term qudit-based quantum processors. The numerical agreement with exact diagonalization establishes a quantitative foundation for scaling such simulations to larger systems where exact methods become intractable.

\section{Conclusion}\label{Sec:Conclusion}
In this work, we developed and analyzed a framework for simulating the dynamics of the $q$-state Potts model on qudit-based quantum architectures using the Suzuki--Trotter decomposition. The central idea is to map the continuous-time evolution under the many-body Hamiltonian onto a sequence of experimentally realizable qudit-native gates. We provided explicit decompositions for both single- and two-qudit unitaries, including a formulation based on the light-shift (LS) interaction and an alternative construction employing Mølmer–Sørensen gates and an auxiliary level to emulate projectors onto the logical subspace.

The numerical simulations for a three-level Potts chain demonstrated that the first-order Suzuki--Trotter approximation accurately reproduces the dynamical quantum phase transition observed in the exact unitary dynamics. The characteristic nonanalyticities in the rate function of the Loschmidt echo were clearly resolved, confirming that the qudit-based decomposition retains the essential non-equilibrium physics of the model. 

These results establish a practical route for implementing multi-level quantum spin models on near-term qudit platforms, such as trapped ions or neutral atoms, where natural access to high-dimensional Hilbert spaces enables compact and hardware-efficient circuit realizations. The presented decompositions and their validation through exact numerical benchmarks form a foundation for extending digital simulation schemes to larger systems and higher-dimensional Potts models, bridging theoretical quantum many-body physics and experimental realizations of multi-level quantum dynamics.

\section*{Author Contributions} 
Conceptualization, M.A.G.; Software, M.A.G.; Investigation, M.A.G., E.O.K.; 
Methodology, E.O.K.;
Writing -- original draft, M.A.G.; Writing -- review \& editing,
E.O.K. and A.S.N.; 
Visualization, A.S.N., M.A.G.; Supervision, A.K.F.; Funding acquisition, A.S.N. and A.K.F. 

All authors have read
and agreed to the published version of the manuscript.

\section*{Institutional Review Board Statement} Not applicable.
\section*{Data Availability Statement} Data are contained within the article.
\section*{Conflicts of Interest} The authors declare no conflicts of interest.

\section*{Funding}
This work of M.A.G., E.O.K. and A.K.F. is supported by the Priority 2030 program at the National University of Science and Technology ``MISIS'' under Project No. K1-2022-027 (generalization of the Potts model, Sections II, III, and V).
The work of A.S.N is supported by RSF Grant No.~24-71-00084 (implementation of composite quantum operations on qudit quantum computing platforms; Section IV).

\section*{Acknowledgments}
The authors thank I.V. Zalivako, A.S. Borisenko, P.A. Kamenskikh and N.V. Semenin for fruitful discussions and useful comments.

\bibliography{bibliography.bib, our_qudits.bib}

\end{document}